\documentclass[11pt,nofootinbib]{revtex4}
\usepackage{amssymb}
\usepackage{amsmath}
\usepackage{graphicx}

\newcommand{\comment}[1]{}

\begin{document}
\baselineskip7mm
\title{A note on differences between (4+1)- and (5+1)-dimensional anisotropic cosmology in the presence 
of the Gauss-Bonnet term}

\author{S.A. Pavluchenko}
\affiliation{Special Astrophysical Observatory, Russian Academy of Sciences, Nizhnij Arkhyz, 369167, Russia}

\author{A.V. Toporensky}
\affiliation{Sternberg Astronomical Institute, Moscow State University, Moscow, 119992, Russia}

\begin{abstract}
We investigate a flat anisotropic (5+1)-dimensional cosmological model in the presence of the Gauss-Bonnet (GB) 
contribution in addition to usual Einstein term in the action. We compared it with (4+1)-dimensional case 
and found a substantial difference in corresponding cosmological dynamics. This difference is manifested in the probability
of the model to have smooth  transition from GB-term-dominated to Einstein-term-dominated phases---this probability
in a reasonable measure on the initial condition space is almost zero for (4+1) case and about 60$\%$ for (5+1) case.
 We discuss this 
difference as well as some features of the dynamics of the considered model.
\end{abstract}

\maketitle


The Lovelock gravity \cite{Lovelock} have been intensively studied last several
years (see, for example~\cite{moreGB,Maeda1,Maeda2.Maeda3,Ferraro,Barrau,Maeda4,Cai,Giribet,D1,D2,Paddy}) showing a revival of interest to this theory after about
a decade of rather slow progress (for achievements of 80-th and early 90-th of the
last century see, for example, \cite{evenmoreGB,N1,N2,Kitaura}). The main feature of this extension of the General Relativity 
is that the resulting equations of motion are  second order differential equations, in contrast to
4-th order equations in other modified gravity theories.
Another important feature of Lovelock gravity is finite number
of terms in the Lagrangian in contrast to string gravity, where there is an infinite
row of additional terms. This means that in perturbative string gravity we can study effects arising
from corrections to GR only at the level of small perturbations, otherwise it would be
necessary to include all infinite set of terms into analysis. In Lovelock gravity 
it is possible to study regimes with non-Einstein terms being equally or more
important than the Einstein part of the action.      

The Lagrangian of Lovelock gravity in an $N$-dimensional space consists of all 
terms which are topological invariants in lower dimensional spaces.
 As a result, there are no corrections to Einstein
theory in ($3+1$) dimensions: the second curvature invariant, Gauss-Bonnet term, appears
in four dimensions and have a contribution to the equations of motion for $N>4$.
The third invariant, cubic in curvature, starts to contribute from ($6+1$) dimensions etc. 
In the present paper we study a flat anisotropic ($5+1$)-dimensional Universe, so the Lovelock Lagrangian
has only one additional term in the form of the Gauss-Bonnet combination. 

In the multidimensional anisotropic cosmology the GB-term leads to two qualitatively new features.
A new type of singularity with finite scale factors and energy density and diverging time derivatives
appears \cite{NS,NS2,NS3}, and recollaps of a flat Universe becomes possible \cite{rc}. In the recent paper \cite{s_ind} it was shown
that in the dynamics of a ($4+1$)-dimensional anisotropic Universe these two regimes are in some sense typical, and in most
cases they prevent the Universe from reaching a low-energy regime from a standard Big Bang singularity.
 Only special initial conditions with
three equal Hubble parameters allow smooth evolution from Big Bang singularity to a 4-dimensional Kasner
regime (which is exact solution in Einstein gravity), all other trajectories either describe recollapsing Universe or meet the nonstandard singularity.

In the present paper we provide a similar analysis for a flat ($5+1$)-dimensional Universe in Lovelock gravity.
We show that the former result is valid only for that particular number of space-time dimensions, and
though in the ($5+1$)-dimensional case there are no additional terms in Lovelock Lagrangian in comparison
with the ($4+1$)-dimensional one, the smooth evolution from high-energy to low-energy regimes in ($5+1$)-dimensional
Universe is typical and does not require severe fine-tuning of the initial conditions.

We consider a flat anisotropic Universe with the metrics 
$ g_{ik}=diag(-1,a^2,b^2,c^2,d^2,e^2)$ and the action
$$
S=\int \sqrt{-g} (R+\alpha GB) d^6 x,
$$
where the Gauss-Bonnet term 
$$
GB=R^{iklm}R_{iklm} -4 R^{ik}R_{ik}+R^2.
$$
Introducing five Hubble parameters in a standard way,
it is 
possible to write down equations of motion in the form of first integral

\begin{equation}
\begin{array}{l}
2H_aH_b+2H_aH_c+2H_aH_d+2H_aH_f+2H_bH_c+2H_bH_d+2H_bH_f+2H_cH_d\\
\\+2H_cH_f+2H_dH_f+24\alpha\left[ \right.
H_aH_bH_cH_d+H_aH_bH_cH_f+H_aH_bH_dH_f\\
\\ \left. +H_aH_cH_dH_f+H_bH_cH_dH_f \right] =0,
 \end{array}
 \label{1}
\end{equation}

\noindent and five dynamical equations. The first equation of motion has the form

\begin{equation}
\begin{array}{l}
2(\dot H_b+H_b^2)+2(\dot H_c+H_c^2)+2(\dot H_d+H_d^2)+2(\dot H_f+H_f^2)+
2H_bH_c+2H_bH_d\\ \\
+2H_bH_f+2H_cH_d+2H_cH_f+2H_dH_f + 
+ 8\alpha \left[3H_bH_cH_dH_f+ (\dot H_b+H_b^2) \right.\\
\\ \times (H_cH_d+H_cH_f+H_dH_f) 
+(\dot H_c+H_c^2)(H_bH_d 
+H_bH_f+H_dH_f)\\ \\  \left.
+(\dot H_d+H_d^2)
(H_bH_c+H_bH_f+H_cH_f)+(\dot H_f+H_f^2)(H_bH_c
+H_bH_d+H_cH_d) \right] =0.
\end{array}
 \label{2}
\end{equation}

\noindent four other equations can be obtained by cyclic transmutation of indices \cite{TT}.

Two obvious limiting cases can be identified easily. First, in the low-curvature regime,
when the contribution from the Gauss-Bonnet term is small, we have well-known generalized
Kasner solution. In this solution scale factors have power-law dependence on time with the power indexes $p_i$,
and, correspondingly, $H_i=p_i/t$ with the Kasner conditions
 $\sum p_i =1$ and $\sum p_i^2=1$.

On the other hand, when Einstein part of equations of motion are negligible in comparison with
the Gauss-Bonnet contribution, another form of power-law solution exists \cite{TT}. There are two form
of this solution, both require $\sum\limits_{i > j > k > 
l} p_i p_j p_k p_l =0$, and either  $\sum\limits_i p_i = 3$ or
$\sum\limits_{i > j > k}p_i p_j p_k =0$.  
The latter case contains solutions with $p_i=(a, b, 0,0,0)$ where $a$ and $b$ are arbitrary numbers.
In addition to discussion in \cite{TT} we can show now that this form is the only possible solution
of the second class up to transmutation of power indices.

Indeed, 
expressing $p_1$ from $\sum\limits_{i > j > k} 
p_i p_j p_k =0$, then substituting it into $\sum\limits_{i > j > k > l} p_i p_j p_k p_l =0$ and expressing $p_2$ we get

\begin{equation}
p_2=-\frac{1}{2} \frac{p_4 p_5 + p_3 p_4 + p_3 p_5 \pm p_3 p_4 p_5 \sqrt{-3p_4^2p_5^2 - 2p_3p_4^2p_5 
-2p_3p_5^2p_4-3p_3^2p_4^2-2p_3^2p_4p_5-3p_3^2p_5^2}}
{p_3 p_4^2 p_5 + p_3^2 p_4^2 + p_3^2 p_4 p_5 + p_3 p_5^2 p_4 + p_3^2 p_5^2 + p_4^2 p_5^2}.
\label{roots}
\end{equation}

\noindent We want to show that the expression under the radical is non-positive. To show this consider 
the case when this expression crosses zero:

\begin{equation}
{\rm Det} = - (3p_4^2p_5^2 + 2p_3p_4^2p_5 +2p_3p_5^2p_4+3p_3^2p_4^2+2p_3^2p_4p_5+3p_3^2p_5^2)=0
\label{det1}
\end{equation}

\noindent It would happen when

\begin{equation}
p_3=- \frac{p_4 p_5 (p_4 + p_5 \pm 2\sqrt{-2p_4^2 - p_4p_5 -2p_5^2})}{3p_4^2+2p_4 p_5 + 3p_5^2}.
\label{roots2}
\end{equation}

\noindent One can easily verify that radical expression in~(\ref{roots2}) is negative for any nonzero $p_4$ and $p_5$.
This means that the expression under radical in (3) has the same sign for any combination of non-zero $p_i$,
an easy check shows that this sign is negative, ruling out all these combinations. If one of $p_i$ is zero,
the condition $\sum\limits_{i > j > k > 
l} p_i p_j p_k p_l =0$
  requires that at least one of remaining indexes is also equal to zero, and in this case the condition
$\sum\limits_{i > j > k}p_i p_j p_k =0$
further requires that the third index also vanishes. As a result, the only possible form 
of the studied class  is the combination $(a,b,0,0,0)$ found in \cite{TT}.

\begin{figure}
\includegraphics[width=0.8\textwidth,bb=0 0 574 862,clip]{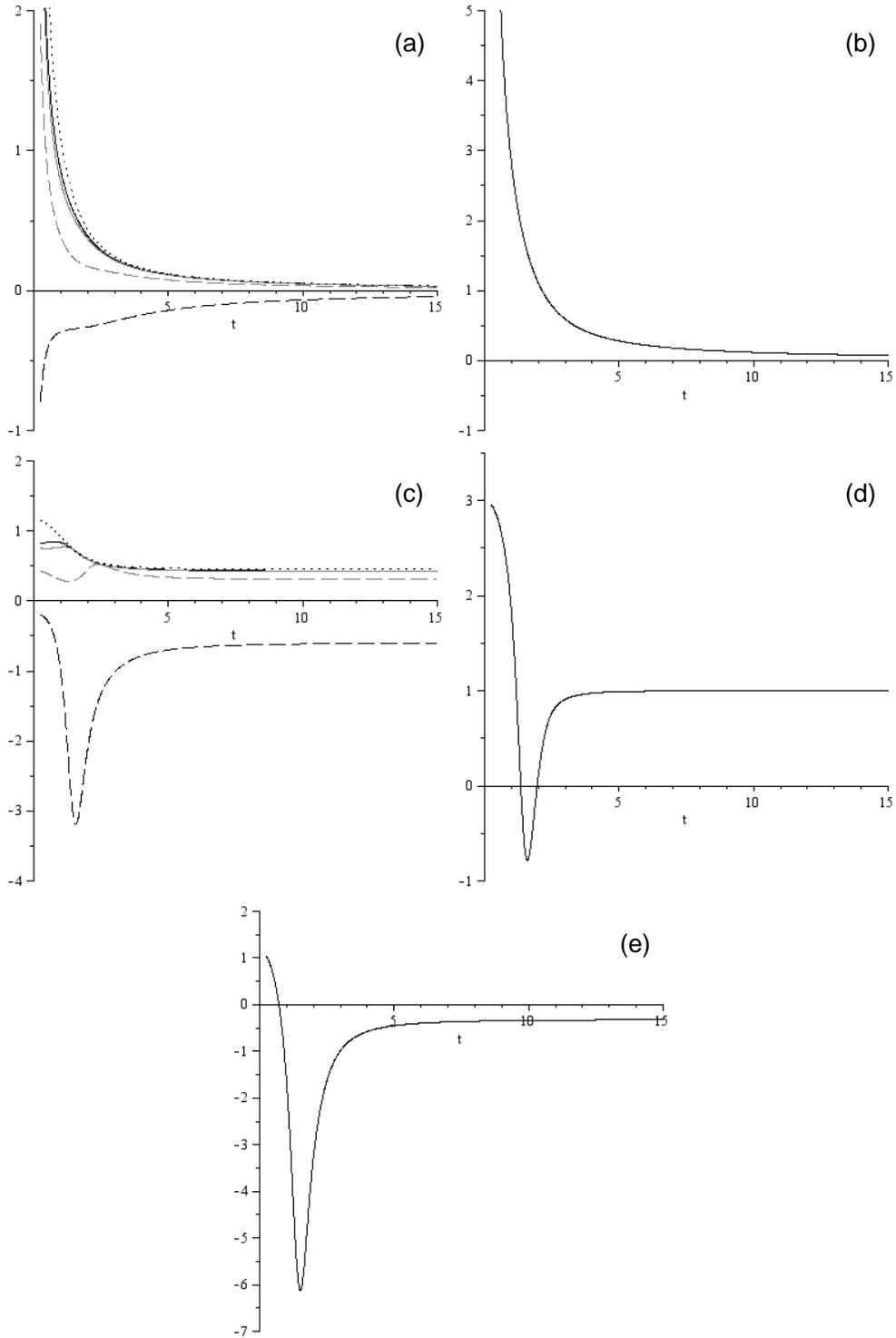}
\caption{Typical behavior of the model with smooth high-to-low-energy power-law
regimes transition. 
In (a) we present the behavior of individual $H_i(t)$,
in (b)---the expansion rate $\sum\limits_i H_i(t)$, in (c)---individual Kasner 
exponents $p_i(t)$,in (d) --- the sum $\sum\limits_i p_i(t)$, and in (e) --- the sum $\sum\limits_{i > j > k} p_i p_j p_k$.}
\end{figure}

This means that this class of solutions is a rather special one, and it is not surprising that
we do not see it in our numerical simulations. From now on we consider only first class of
power-law solution as a solution with standard cosmological singularity in a high-energy regime.

Apart from high-curvature and low-curvature asymptotic power-law regimes the system (1--2) describes 
complicated behavior in the range of Hubble parameters where both Gauss-Bonnet and Einstein contributions
are important. We have studied this system numerically, starting from initial conditions distributed 
randomly in the interval [0,1] for three Hubble parameters, in the interval [-1,1] for one of them  
and calculating fifth Hubble parameter from
the constraint equation~(1). The integration stopped when one of the following regimes are reached:
\begin{itemize}
\item For expanding Universe: low-curvature Kasner regime, nonstandard singularity or recollaps.
\item For contracting Universe: high-curvature power-law regime ("standard singularity") or nonstandard
singularity.
\end{itemize}
Matching two branches, expanding and contracting, of numerical solution allows us to construct a full history
of the modeled Universe evolving from one of listed states to another.

The main feature which distinguishes the (5+1) case from the (4+1) one is that now the transition from a standard
singularity to low-curvature Kasner regime does not require severe fine-tuning (we remind a reader that in the
former case only trajectories with three scale factors equal to each other can show this type of transition).

In Fig.~1 we presented a typical behavior of the model with ``standard singularity'' $\to$ Kasner
smooth transition. In (a) panel we show $H_i(t)$ evolution curves, in (b) panel---$\sum\limits_i H_i(t)$;
individual Kasner exponents $p_i$ are shown in (c) and their sum $\sum\limits_i p_i(t)$ in (d). One can easily
see that this is a high-energy power-law ($\sum\limits_i p_i(t) = 3$) to low-energy Kasner
($\sum\limits_i p_i(t) = 1$) transition indeed. We have found that about 60$\%$ of all the trajectories belong to
this class.
Studying examples of this kind of dynamics for various initial conditions we also noticed that final values
of power indexes are not distributed in homogeneous way on the Kasner sphere, but prefer the situation clearly
seen in the Fig.1(c) with one negative index and four positive and close to each other.
Inserting $p_1 \approx p_2 \approx p_3 \approx p_4$ into 
the Kasner conditions one can found that $\sum\limits_{i > j > k}p_i p_j p_k \approx -0.3$. This estimation meets the
results of our numerical analysis fairy well, which can be seen from Fig.1(e) where the combination
$\sum\limits_{i > j > k}p_i p_j p_k$ is plotted.
The reason for $p_i$ to be approximately equal in the
low-energy stage remains a mystery and may indicates that some hidden symmetries we are ignorant about are
involved, so the problem requires more detailed investigation.

\begin{figure}
\includegraphics[width=0.85\textwidth,bb=0 0 574 285,clip]{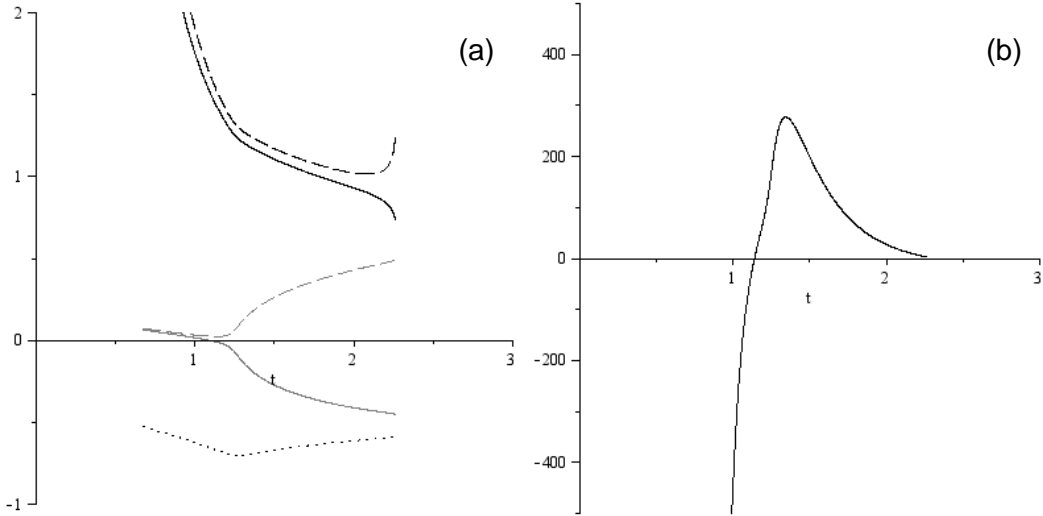}
\caption{Typical behavior of the model which experiences a non-standard singularity.
In panel (a) we presented  Hubble functions $H_i(t)$ and in (b)---the denominator of $\dot H_i$.}
\end{figure}

One of ``non-smooth'' solutions indicated above is the case of nonstandard singularity. This is the case when sometime
during the evolution $\dot H_i$ diverges while $H_i$ remain finite.
Solving eqs.(2) with respect to $\dot H_i$ we can express them in the form
of fraction, and the nonstandard singularity occurs when the denominator crosses zero while numerator is regular 
and non-zero.
In Fig.~2 the behavior of our model is shown for this case.  In panel (a) we present the
Hubble functions $H_i(t)$,  
in panel (b) --- the behavior of the denominator of the fractional expression for $\dot H_i$.
 At $t \approx 1.2$ the denominator crosses zero, numerator and Hubble parameters remain regular and non-zero.
 As this singularity is weak according to the Tipler terminology \cite{Tipler}, our numerical program typically goes through it,
 and the numerical results after the singularity are not physically valid: despite  $H_i(t)$ are regular,
 $\dot H_i(t)$ are singular, and, hence, scalar curvature invariants diverge,
  which makes the non-standard singularity physical, but not  a coordinate one.
 So, the evolution
presented in Fig.~2\,(a) is only valid  till the point $t \approx 1.2$ where 
the non-standard singularity occurs. Depending on initial conditions, the nonstandard singularity
 can be found either in future or past history of the particular
Universe. These trajectories take up to 15$\%$ of all the trajectories. 

 \begin{figure}
\includegraphics[width=0.8\textwidth,bb=0 0 574 862,clip]{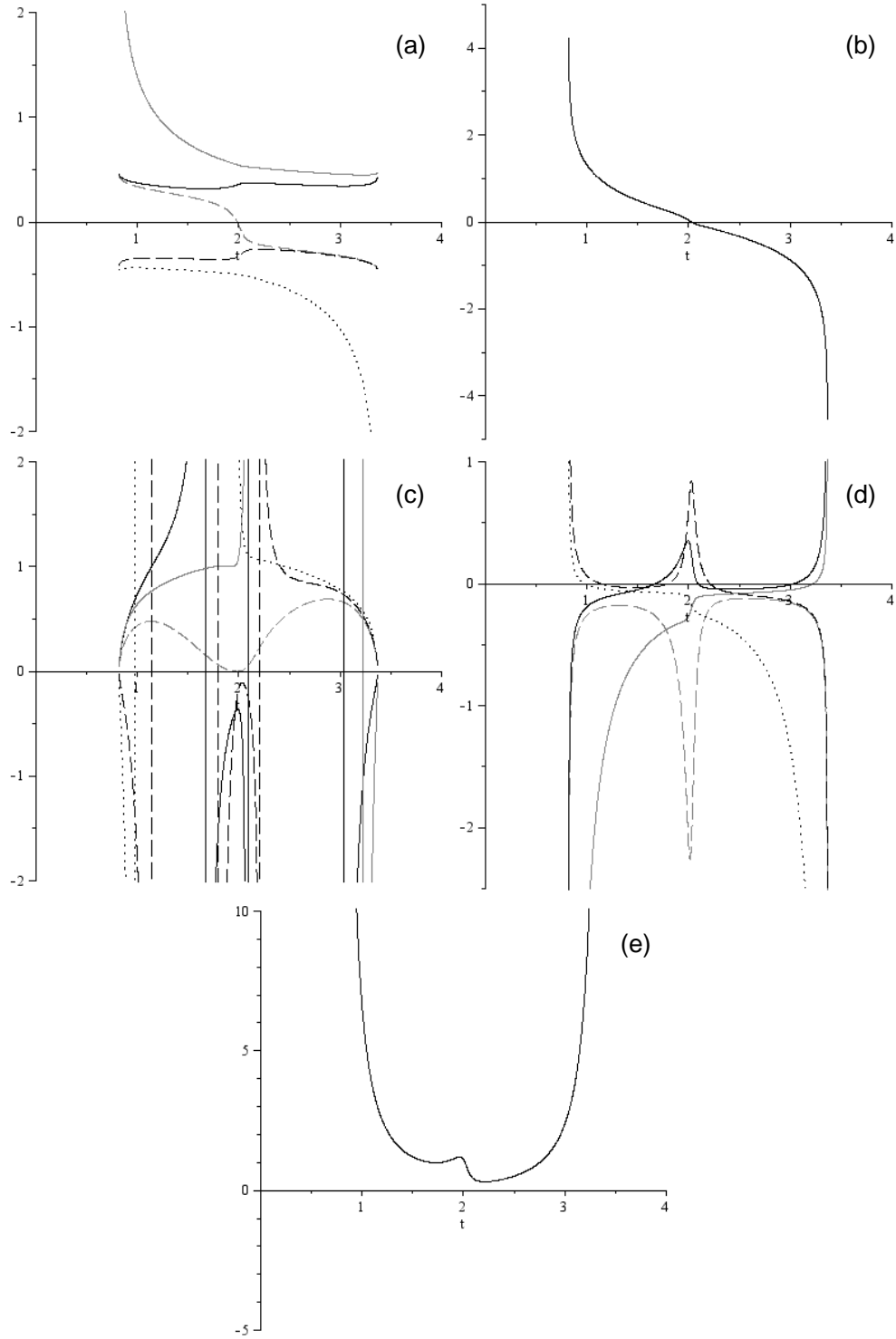}
\caption{Typical behavior of the model which experiences a recollaps.
In (a) we present the behavior of  $H_i(t)$,
in (b)--- the expansion rate $\sum\limits_i H_i(t)$, in (c)---the Kasner 
exponents $p_i(t)$; $\dot H_i(t)$ in (d) and their denominator in (e).}
\end{figure}

Finally, the last case is the recollaps. In this case,  starting from  expansion (the
initial conditions in this case are chosen in a way to ensure $\sum\limits_i H_i > 0$ from the beginning),
the Universe begins to contract at 
some point and ends up in a singularity.
 In Fig.~3 we presented a typical behavior of the model which experience recollaps.  In the panel (a) 
we showed individual $H_i(t)$, their sum in (b), individual $p_i(t)$ in (c).
They diverge at some points
due to the definition of $p_i$: $p_i = - H_i^2 / \dot H_i$.  In the panel (d) we presented $\dot H_i(t)$ and one can verify
zeros of $\dot H_i(t)$ correspond to ``singularities'' of $p_i(t)$.
Also in panel (e) of the Fig.~3
we presented the denominator of the $\dot H$, so the reader can verify that
singularities in $p_i$  have nothing to do with the nonstandard singularity described above. Last thing to mention, this case occupy remaining 25$\%$ of all the
trajectories.

We have considered dynamics of a flat anisotropic vacuum ($5+1$)-dimensional Universe in Gauss-Bonnet gravity.
Unlike the ($4+1$)-dimensional case studied earlier~\cite{s_ind}, we have found that nonstandard singularity and possible
recollaps which usually prevented five-dimensional Universe from reaching a low-curvature regime 
do not dominate in six dimensions. About 60\% of trajectories in the latter case show
smooth transition from GB dominated epoch to a low-curvature six-dimensional Kasner regime. 
The other possibilities are 
a ``non-standard'' singularity (about 15\%) and a recollaps (about 25\%). As we have used random set of initial
conditions, other realizations can show slightly different numbers. However, our main result has a qualitative 
character and could not change significantly --- a transition from Big Bang to low-curvature asymptotic is rather general
in ($5+1$) dimensions in contrast to the ($4+1$)-dimensional case. 
This may indicate 
that Lovelock cosmology in the world with odd number of space dimensions is much less pathological than in 
even-dimensional worlds. 

Some dynamical features of the model still require additional investigation. Another possible direction of future
work is to include a matter into analysis, which can modify the results for both ($4+1$)- and ($5+1$)-dimensional cases.

\section*{Acknowledgments}
This work is partially supported by RFBR grant 08-02-00923 and scientific school grant 4899.2008.2.

\end{document}